\documentstyle[12pt]{article}

\begin{document}

\author{Hao Chen \\
Department of Mathematics\\
Zhongshan University\\
Guangzhou,Guangdong 510275\\
People's Republic of China}
\title{Algebraic-geometric separability criterion and low rank mixed state entanglement}
\date{June,2002}
\maketitle

\begin{abstract}
We first propose a new separability criterion based on algebraic-geometric invariants of bipartite mixed states introduced in [1], then prove that for all low ranks $r \leq m+n-3$, generic rank $r$ mixed states in $H_A^m \otimes H_B^m$ have realtively high Schmidt numbers by this separability criterion (thus entangled). This also means that the algebraic-geometric separability criterion prposed here can be used to detect all low rank entagled mixed states outside a measure zero set.
\end{abstract}

Quantum entanglement was first noted as a feature of quantum mechanics in
the famous Einstein, Podolsky and Rosen [2] and Schrodinger [3] papers. Its
importance lies not only in philosophical considerations of the nature of
quantum theory, but also in applications where it has emerged recently that
quantum entanglement is the key ingredient in quantum computation [4] and
communication [5] and plays an important role in cryptography [6,7].\\

A mixed state $\rho$ in the bipartite quantum system $H=H_A^m \otimes H_B^n$
is called separable if it can be written in the form $\rho=\Sigma_j p_j
|\psi_j><\psi_j| \otimes |\phi_j><\phi_j|$, where $p_j >0$ and $%
|\psi_j>,|\phi_j>$ are pure states in $H_A^m,H_B^n$. Otherwise it is called
entangled, ie., it cannot be prepared by A and B separately. It is realized
that entangled states are very important resources in quantum communication,
quantum cryptography and quantum computation.  \\

To find good necessary conditions of separability (separability criteria) is one of the fundamental problem in the study of quantum entanglement (see [8],[9]). Bell's inequality and entropy criterion ([8],[9]) are well-known scalar criterion. Recent disorder criterion [10] can be thought as a stronger form of previously known entropy criteria. In 1996 A.Peres proposed a striking simple separability criterion ([11]) which asserts that a separable mixed state $\rho$ necessarily has (semi)-positive partial transpose (PPT, ie., the partial transpose $\rho^{PT}$, $<ij|\rho^{PT}|kl>=<il|\rho|kj>$, has  no negative eigenvalue). There is another criterion called reduction criterion ([12]) which is weaker than PPT criterion. In [13] it is proved that for "very low" rank mixed states $\rho$ in $H_A^m \otimes H_B^n$ with rank $\rho \leq max \{m,n\}$, PPT is equivalent to the separability.\\

Among other things, the so-called range criterion proposed in [14] played a very important role in constructing PPT entangled mixed states. In [14],[15], [16],[17],[18],[19] many (families) of PPT entangled mixed states are constructed by proving that there is no separable pure state in their ranges, and thus violated the range criterion and entangled. Quite recently some new separability criteria were proved ([20],[21],[22])and they are strong in the sence that they can be used to detect most previously known PPT entangled mixed states. We introduced algebraic-geometric invariants of bipartite mixed states as their non-local invariants and proved a new separability criterion in [1], by which some PPT entangled mixed states whose ranges contain some separable pure states were constructed.\\

Upto now it seems that people cannot expect a separability criterion which can detect all (or almost all=generic=outside a measure zero set) entangled mixed states. In this aspect the only known result is [13]. It is proved that for bipartite mixed states $\rho$ in $H_A^m \otimes H_B^n$ with ranks not bigger than $max \{m,n\}$ the PPT property is equivalent to  the separability, thus PPT criterion can be used to detect all "very low" rank entangled bipartite mixed states. However it is well-known that there exists rank $max\{m,n\}+1$ entangled PPT bipartite mixed states in $H_A^m \otimes H_B^n$ ([18],[1]), thus PPT criterion cannot dectect entangled mixed states with ranks bigger than $max\{m,n\}$.\\

In this letter we propose a separability criterion based on algebraic-geometric invariants of bipartite mixed states introduced in [1] and prove by this criterion that generic $rank (\rho) \leq m+n-3$ bipartite mixed states $\rho$ in $H_A^m \otimes H_B^n$ are entangled. Thus we can see that this criterion can detect almost all low rank bipartite enatngled mixed states.\\

In [1] we introduced a series of algebriac sets $V_A^k(\rho)$ in $CP^{m-1}$ (resp. $V_B^k(\rho)$ in $CP^{n-1}$) , where$k=0,1,...$, (ie., zero locus of some homogeneous multivariable polynomials, actually these polynomials are the determinants of some matrices, see[23]for mathematics of algebraic sets) as invariants of the bipartite mixed states in $H_A^m \otimes H_B^n$ under local unitary transformations. It is also proved in [1] that these algebraic-geometric invariants are the sum of some linear subspaces of $CP^{m-1}$ (resp. $CP^{n-1}$)if the mixed states are separable. This is the 1st algebraic-geometric separability criterion. The 2nd algebraic-geometric separability criterion proposed here is a lower bound of Schmidt numbers (see [24]) based on the emptyness of some algebraic-geometric invariants of the mixed states. Let us first recall the definition of Schmidt numbers of mixed states in [24]. For a bipartite mixed state $\rho$, it has Schmidt number $k$ if and only if for any decomposition $\rho=\Sigma_i p_i |v_i><v_i|$ for positive real
numbers $p_i$'s and pure states $|v_i>$'s, at least one of the pure states $%
|v_i>$'s has Schmidt rank at least $k$, and there exists such a
decomposition with all pure states $|v_i>$'s Schmidt rank at most $k$. It is clear that the mixed states are entangled if their Schmidt numbers are bigger than 1. It is
proved ([24]) that Schmidt number is entanglement monotone, ie., they cannot
increase under local quantum operations and classical communication .
So we can naturally think Schmidt numbers of mixed states as a measure of
their entanglement. \\

The main results of this letter are the following results.\\

{\bf Theorem 1 ( separability criterion).} {\em Let $\rho$ be a mixed state on $H_A^m \otimes H_B^m$ of rank $r$ and Schmidt number $k$.\\ 
1) If $V_A^{m-t}(\rho)=\emptyset$, $k \geq \frac{m}{r-m+t}$;\\
2) If $V_B^{n-t}(\rho)=\emptyset$, $k \geq \frac{n}{r-n+t}$.}\\

From this lower bound of Schmidt numbers of bipartite mixed states we have\\

{\bf Theorem 2.} {\em There exists a subset $Z(r)$ defined by algebraic
equations in the space $M(r)$ of all rank $r$ mixed states in bipartite
quantum system $H_A^m \otimes H_B^m$ , such that, Schmidt numbers of mixed
states in $M(r)\setminus Z(r)$ are at least\\
1) $[\sqrt{m}]-1$ if $r\leq m$ ($[\sqrt{n}]-1$ if $r\leq n$);\\
2) $\frac{m}{r-m+[\sqrt{m}]+1}$ if $r>m$ ($\frac{n}{r-n+[\sqrt{n}]+1}$ if $r>n$ );\\
3) $3$ if $\frac{3m}{2}-5 \geq r >m \geq 169$ ($3$ if $\frac{3n}{2}-5 \geq r >n \geq 169$);\\
4) $2$ if $r \leq m+n-3$.}\\

Here $[x]$ of a positive real number $x$ means the integral part of $x$.\\

We should note that the algebraic set $Z(r)$  defined as zero locus of
algebraic equations has volume zero under any reasonable measure. Thus it is
known from Theorem 1 that random low rank mixed states in $H_A^m \otimes
H_B^m$ are highly entangled, for example, random rank $r \leq m+\sqrt{m}$
mixed states have their Schmidt numbers at least $[\sqrt{m}]/2-1$ from 1)
and 2) of Theorem 1.\\

For example, for rank 1 mixed states (pure states) $\phi=\Sigma_{i,j}^m
a_{ij} |ij>$ it is well-known that the Schmidt number of $\phi$ is just the
rank of the matrix $A=(a_{ij})_{1\leq i,j \leq m}$. Thus the pure states
outside the algebraic set defined by $det A=0$ have Schmidt number $m$, thus
highly entagled. This is previously known result ([25],[26]).\\

We shuld indicate how to use the separability criterion Theorem 1 for a given concrete bipartite mixed states. Let us recall the definition of algebraic-geometric invaraints in [1]. For any bipartite mixed states $\rho$ on $H_A^m \otimes H_B^n$ , we want to understand it by measuring it with separable pure states, ie., we consider the $<\phi_1 \otimes \phi_2
|\rho|\phi_1 \otimes \phi_2>$ for any pure states $\phi_1 \in H_A^m$ and $\phi_2 \in H_B^n$. For any fixed $\phi_1 \in P(H_A^m)$, where $P(H_A^m)$ is
the projective space of all pure states in $H_A^m$, $<\phi_1 \otimes \phi_2
|\rho|\phi_1 \otimes \phi_2>$ is a Hermitian bilinear form on $H_B^n$,
denoted by $<\phi_1|\rho|\phi_1>$ . We consider the {\em degenerating locus }
of this bilinear form, ie., $V_A^k(\rho)=\{\phi_1 \in P(H_A^m): rank
(<\phi_1|\rho|\phi_1>) \leq k\}$ for $k=0,1,...,n-1$. We can use the
coordinate form of this formalism. Let $\{|11>,...,|1n>,...,|m1>,...,|mn>\}$
be the standard orthogonal base of $h_A^m \otimes H_B^n$ and $\rho$ be an
arbitrary mixed states. We represent the matrix of $\rho$ in the base $\{|11>,...|1n>,...,|m1>,...,|mn>\}$, and consider $\rho$ as a blocked matrix 
$\rho=(\rho_{ij})_{1 \leq i \leq m, 1 \leq j \leq m}$ with each block $\rho_{ij}$ a $n \times n$ matrix corresponding to the $|i1>,...,|in>$ rows
and the $|j1>,...,|jn>$ columns. For any pure state 
$\phi_1=r_1|1>+...+r_m|m> \in P(H_A^m)$ the matrix of the Hermitian linear
form $<\phi_1|\rho|\phi_1>$ with the base $|1>,...,|n>$ is $\Sigma_{i,j}
r_ir_j^{*} \rho_{ij}$. Thus the ``degenerating locus'' is actually as
follows.\\

$$
\begin{array}{ccccc}
V_{A}^k(\rho)=\{(r_1,...,r_m)\in CP^{m-1}:rank( \Sigma_{i,j}r_ir_j^{*}
\rho_{ij}) \leq k\} &  &  &  & 
\end{array}
$$
for $k=0,1,...,n-1$. Similarly $V_{B}^k (\rho) \subseteq CP^{n-1}$ can be
defined. Here * means the conjugate of complex numbers. It is known from
Theorem 1 and 2 of [1] that these sets are algebraic sets (zero locus of
several multi-variable polynomials, see [21]) and they are invariants under
local unitary transformations depending only on the eigenvectors of $\rho$.
Actually these algebraic sets can be computed easily as follows.\\

Let $\{|11>,...,|1n>,...,|m1>,...,|mn>\}$ be the standard orthogonal base of 
$H_A^m \otimes H_B^n$ as above and $\rho= \Sigma_{l=1}^{t} p_l |v_l><v_l|$
be any given representation of $\rho$ as a convex combination of projections
with $p_1,...,p_t >0$. Suppose $v_l=\Sigma_{i,j=1}^{m,n} a_{ijl} |ij>$ , $A=(a_{ijl})_{1\leq i \leq m, 1 \leq j \leq n, 1 \leq l \leq t}$ is the $mn
\times t$ matrix. Then it is clear that the matrix representation of $\rho$
with the base $\{|11>,...,|1n>,...,|m1>,...,|mn>\}$ is $AP(A^{*})^{\tau}$,
where $P$ is the diagonal matrix with diagonal entries $p_1,...,p_t$. We may
consider the $mn\times t$ matrix $A$ as a $m\times 1$ blocked matrix with
each block $A_w$, where $w=1,...,m$, a $n\times t$ matrix corresponding to 
$\{|w1>,...,|wn>\}$. Then $V_A^k(\rho)$ is just the algebraic set in $CP^{m-1}
$ as the zero locus of the determinants of all $(k+1) \times (k+1)$
submatrices of $\Sigma_i^m r_i A_i$.\\

It is clear that $V_A^k(\rho)$ contains $V_A^{k-1}(\rho)$ and it is more difficult to compute $V_A^h(\rho)$ for smaller $h$. In the real application of Theorem 1, we find the biggest $h$ such that $V_A^h(\rho)=\emptyset$ and from Theorem 1 it is known that the Schmidt number of $\rho$ is at least $\frac{m}{r-h}$, where $r=rank(\rho)$. From Theorem 2 for generic $r  \leq m+n-3$ mixed states in $H_A^m \otimes H_B^n$ , this proceture implies that $k \geq 2$ and $\rho$ is entangled. We can see that to check if $V_A^h(\rho) =\emptyset$ is equivalent to check if there is any non-zero solution for some homogeneous multivariable polynomials.\\

Theorem 1 can be used to give strong results for Schmidt numbers of mixed
states ,especially it is very useful to dectect low rank
entangled mixed states as illustrated in the following examples.\\

{\bf Example 1.} Let $\rho$ be a rank 2 mixed state in $H_A^m \otimes H_B^m$
of the form $\rho=\lambda_1 |v_1><v_1|+\lambda_2 |v_2><v_2|$, where $%
\lambda_1$ and $\lambda_2$ positive, $v_1$ and $v_2$ are linear independent
unit vectors and of the form $v_1=\Sigma_{ij}a_{ij}^1|ij>$ and $%
v_2=\Sigma_{ij} a_{ij}^2 |ij>$. Suppose the linear span by the $2m$ rows of
the matrices $A^1=(a_{ij}^1)_{1\leq i,j\leq m}$ and $A^2=(a_{ij}^2)_{1\leq
i,j \leq m}$ is of dimension $m$ (We should note that this is a condition
satisfied by generic rank two mixed states, since for all rank 2 mixed
states outside an algebraic set, the $2m \times m$ matrix consisting of $2m$
rows of $A^1$ and $A^2$ has rank $m$). Then Schmidt number of $\rho$ is at
least $\frac{m}{2}$.\newline

For the proof, we just need to take $r=2,t=m$ in Theorem 2. From the
condition that the $2m \times m$ matrix consiting of $2m$ rows of $A^1$ and $%
A^2$ has rank $m$, we can easily get $V_A^0(\rho)=\emptyset$. Thus from
Theorem 2, we get the conclusion.\newline

For example, the following rank 2 mixed states $\rho_{\lambda_1,\lambda_2}=%
\frac{1}{\lambda_1+\lambda_2}(\lambda_1 |v_1><v_2|+ \lambda_2 |v_2><v_2|)$
in bipartite quantum system $H_A^5 \otimes H_B^5$, where $\lambda_1$ and $%
\lambda_2$ are real positive real numbers and \newline
$$
\begin{array}{cccccc}
v_1=\frac{1}{\sqrt{2}}(|11>+|22>) &  &  &  &  &  \\ 
v_2=\frac{1}{2}(|33>+|44>+|55>+|45>) &  &  &  &  & 
\end{array}
$$

have their Schmidt numbers at least 3.\newline

Actually example 1 can be generalized as follows.\newline

{\bf Corollary 1.} {\em Let $\rho=\Sigma_{t=1}^r p_i |v_t><v_t|$ be a rank $%
r < m$ mixed state in $H_A^m \otimes H_B^m$, where $p_1,...,p_r$ are
positive numbers and $v_t=\Sigma_{ij} a_{ij}^t|ij>, A^t=(a_{ij}^t)_{1 \leq
i,j \leq m}$. Suppose that the linear span of all $rm$ rows of matrices $%
A^1,...,A^r$ is of dimension $m$. Then Schmidt number of $\rho$ is at least $%
\frac{m}{r}$,thus entangled.}\newline

{\bf Proof.} We take $t=m$ in Theorem 2 and from the condition about
matrices $A^1,...,A^r$ it is clear $V_A^0(\rho)=\emptyset$. Thus the
conclusion follows from Theorem 2.\newline

Corollay 1 implies that if a mixed state is mixed by not too many pure
states and one of these pure states has highest Schmidt rank, then the mixed
state has a relatively high Schmidt number. It is clear that the condition
of Corollary 1 is satisfied by generic rank $r<m$ mixed states.\newline

{\bf Example 2.} Let $\rho_{\lambda_1,\lambda_2,\lambda_3}=\frac{1}{%
\lambda_1+\lambda_2+\lambda_3}(\lambda_1 |v_1><v_2|+\lambda_2
|v_2><v_2|+\lambda_3 |v_3><v_3|)$ in bipartite quantum system $H_A^7 \otimes
H_B^7$, where $\lambda$'s are positive real numbers, $v_1=\frac{1}{\sqrt{7}}%
(|11>+\cdots+|77>)$ and $v_2$ and $v_3$ are arbitray pure states. Then
Schmidt numbers of $\rho_{\lambda_1,\lambda_2,\lambda_3}$ are at least 3.\\

Generally the follwing result is valid.\newline

{\bf Corollary 2.}{\em Let $\rho=\Sigma_{i=1}^r p_i |v_i><v_i|$ be a mixed
state in $H_A^m \otimes H_B^m$, where $p_1,...,p_r$ are positive real
numbers and Schmidt rank of $v_1$ is $m$. Then Schmidt number of $\rho$ is
at least $\frac{m}{r}$ and thus\\
1) Schmidt number of $\rho$ is at least 3 if $r <\frac{m}{2}$;\newline
2) $\rho$ is entangled when $r<m$.}\\

{\bf Example 3.} Let $\rho=\frac{1}{3}(|\phi_1><\phi_1|+|\phi_2><\phi_2|+|\phi_3><\phi_3|)$ be a rank 3 mixed state in $H_A^3 \otimes H_B^3$, where\\

$$
\begin{array}{cccccccccccc}
\phi_1=\frac{1}{\sqrt{2}}(|11>+|33>)\\
\phi_2=\frac{1}{\sqrt{2}}(|12>+|21>)\\
\phi_3=\frac{1}{\sqrt{3}}(|13>+|22>+|31>)
\end{array}
$$

Then $\Sigma r_i A_i$ is of the folllowing form\\

$$
\left(
\begin{array}{cccccccccc}
r_1&r_2&r_3\\
0&r_1&r_2\\
r_3&0&r_1\\
\end{array}
\right)
$$

It is clear that $rank(\Sigma r_i A_i) \leq 1$ implies $r_1=r_2=r_3=0$, thus $V_A^1(\rho)=\emptyset$. From Theorem 1 we have the Schmidt number of $\rho$ is at least 2 thus $\rho$ is entangled.\\

{\bf Example 4.}  Let $\rho=\frac{1}{5}(|\phi_1><\phi_1|+|\phi_2><\phi_2|+|\phi_3><\phi_3|+|\phi_4><\phi_4|+|\phi_5><\phi_5|)$ be a rank 5 mixed state in $H_A^4 \otimes H_B^4$, where\\

$$
\begin{array}{cccccccccccc}
\phi_1=|11>\\
\phi_2=\frac{1}{\sqrt{2}}(|12>+|21>)\\
\phi_3=\frac{1}{2}(|13>+|22>+|31>+|44>)\\
\phi_4=\frac{1}{\sqrt{3}}(|23>+|32>+|41>)\\
\phi_5=\frac{1}{\sqrt{2}}(|33>+|42))
\end{array}
$$

Then $\Sigma r_i A_i$ is of the folllowing form\\

$$
\left(
\begin{array}{cccccccccc}
r_1&r_2&r_3&r_4&0\\
0&r_1&r_2&r_3&r_4\\
0&0&r_1&r_2&r_3\\
0&0&r_4&0&0\\
\end{array}
\right)
$$

It is clear that $rank(\Sigma r_i A_i) \leq 2$ implies $r_1=r_2=r_3=r_4=0$, thus $V_A^2(\rho)=\emptyset$. From Theorem 1 we have the Schmidt number of $\rho$ is at least 2 thus $\rho$ is entangled.\\

From Example 1,2,3,4 and Corollary 1,2 we can see that our method is
constructive to check Schmidt numbers of mixed states by just calculating
rank of some numerical matrices.\newline

For pure states $\rho=|v><v|$ in $H_A^m \otimes H_B^n$ with $m \leq n$, we can compute its algebraic-geometric invariants  from its Schmidt decomposition $v= \Sigma_{i=1}^d a_i e_i \otimes e'_i$, where $e_1,...,e_m$ (resp., $e'_1,...,e'_n$) is a orthogonal base of $H_A^m$ (resp. $H_B^n$). It is clear that $V_A^0(\rho) =\{(r_1,...,r_m) \in CP^{m-1}: (a_1r_1,...,a_d r_d,0,...,0)^{\tau}=0\}$. Thus we have\\

{\bf Proposition 1.} {\em For the pure state $\rho=|v><v|$, $d=m$ if and only if $V_A^0(\rho)=\emptyset$ and $d=m-1-dim(V_A^0(\rho))$ if $d \leq m-1$.}\\

The following obsevation is the the key point of the proof of Theorem 1.
From Lemma 1 in [14], if $\rho=\Sigma_{i=1}^t p_i |v_i><v_i|$ with positive $p_i$'s
the range of $\rho$ is the linear span of vectors 
$|v_1>,...,|v_t>$. We take any $dim(range(\rho))$ linear independent vectors
in the set $\{|v_1>,...,|v_t>\}$, say they are $|v_1>,...,|v_s>$ , where 
$s=dim(range(\rho))$. Let $B$ be the $mn \times s$ matrix with columns
corresponding to the $s$ vectors $|v_1>,...,|v_s>$'s coordinates in the
standard base of $H_A^m \otimes H_B^n$. We consider $B$ as $m \times 1$
blocked matrix with blocks $B_1,...,B_m$ $n \times s$ matrix as above. It is
clear that $V_A^k(\rho)$ is just the zero locus of determinants of all $%
(k+1) \times (k+1)$ submatrices of $\Sigma_i^m r_iB_i$, since any column in $%
\Sigma_i r_i A_i$ is a linear combination of columns in $\Sigma_i r_i B_i$.\\

{\bf Proof of Theorem 1.} We just prove the coclusion 1). Take a representation $\rho= \Sigma_{i=1}^t p_i |v_i><v_i|$
with $p_i$'s positive, and the maximal Schmidt rank of $v_i$'s is $k$. As
observed above, it only need to take $r$ linear independent vectors in $%
\{v_1,...,v_t\}$ to compute the rank of $\Sigma_i r_i A_i$. For the purpose
that the rank of these $r$ columns in $\Sigma_i r_i A_i$ is not bigger than $%
m-t$, we just need $r-m+t$ of these columns are zero. On the other hand,
from Proposition 1 the dimension of the linear subspace $%
(r_1,...,r_m) \in H_A^m$, such that the corresponding column of $v_i$ in $%
\Sigma_i r_i A_i$ is zero, is exactly $m-k(v_i)$ where $k(v_i)$ is the
Schmidt rank of $v_i$. Thus we know that there is at least one nonzero $%
(r_1,...,r_m)$ such that $\Sigma_i r_i A_i$ is of rank smaller than $m-t+1$
if $m >k(r-m+t)$. The conclusion is proved.\\

For the purpose to prove Theorem 2, we need to recall a well-known result in
the theory of determinantal varieties (see Proposition in p.67 of [27]). Let 
$M(m,n)=\{(x_{ij}): 1\leq i \leq m, 1 \leq j \leq n\}$ (isomorphic to 
$CP^{mn-1}$) be the projective space of all $m \times n$ matrices. For a
integer $0 \leq k \leq min\{m,n\}$, $M(m,n)_k$ is defined as the locus $%
\{A=(x_{ij}) \in M(m,n): rank(A) \leq k\}$. $M(m,n)_k$ is called generic
determinantal varieties.\\

{\bf Proposition 2.} {\em $M(m,n)_k$ is an irreducible algebriac subvariety
of $M(m,n)$ of codimension $(m-k)(n-k)$.}\newline

We describe the basic idea of Proposition 2. Since all entries in the $m\times n$ matrix are indeterminants
, we can suppose that the 1st $k \times k$ submatrix is nonsingular and the 
remaining $m-k$ columns ($n-k$ rows) are linear dependent on the 1st $k$ columns
($k$ rows). This condition implies that the determinants of all $(m-k)(n-k)$ $(k+1) \times (k+1)$ submatrices containing
1st $k \times k$ submatrix are zero, ie., we have $(m-k)(n-k)$ (independent) algebraic equations to define $M(m,n)_k$, thus the conclusion of Proposition 2 is valid.\\  

Now we can prove Theorem 2, the idea is basically the same as the proof of
Corollary 1,2, ie., we take suitable $t$ such that $V_A^{m-t}(\rho)$ has
codimension larger than $m-1$ and then apply Theorem 1 to get the conclusion.\\

{\bf Proof of Theorem 2.} Without loss of generality we assume $m \geq n$ and just prove the conclusions for $m$. Similarly as the argument for Corollary 1.2 for any
given $m,r$ for the purpose that $V_A^{m-t}(\rho)=\emptyset$ for generic
mixed states in $H_A^m \otimes H_B^m$ we just need $(m-(m-t))(r-(m-t)) \geq m
$ from Proposition 2. \\

We take $t=m-r+\sqrt{m}+1$ in case 1) and $t=\sqrt{m}+1$ in case 2) , the
conclusions of 1) and 2) are proved.\\

It is clear that generic $r \leq \frac{4m-3-3[\sqrt{m}]}{3}$ has their
Schmidt numbers at least $3$ from 1) and 2). For ranks $\frac{4m-3-3[\sqrt{m}]}
{3} \leq r \leq \frac{3m}{2}-5$ we can take $t=4$, the conclusion of 3) is
proved.\\

For mixed states of rank $r=m+n-3$, we take $t=m-n+2$. A similar argument as
above implies that $V_A^{n-2}(\rho)$ in $CP^{m-1}$ has codimension 
$(m-n+2)(m-1)>m-1$, thus empty for generic $m+n-3$ mixed states. Hence from Theorem
1, Schmidt numbers of generic rank $m+n-3$ mixed states are at least 
$\frac{m}{m+n-3-n+2}=\frac{m}{m-1}$. For other ranks we can use a similar argument to
get the conclusion.\\

Since algebraic-geometric invariants $V_A^k(\rho)$ of mixed states $\rho$
are only dependent on the range of $\rho$. Thus our results actually imply
that generic low dimension $r \leq 2m-3$ subspaces of $H_A^m \otimes H_B^m$
cannot be a linear span of separable pure states(recall range criterion of
P. Horodecki in [20]). However this is not true for high dimensional
subspaces of $H_A^m \otimes H_B^m$. In the following example it is proved
that generic dimension 3 subspaces of $H_A^2 \otimes H_B^2$ are linear span
of separable pure states.\\

{\bf Example 5.} For any given 4 complex numbers $a,b,c,d$ satisfying $d\neq
0 $ and $ad \neq bc$, it is easy to check that the following $3$ product
vectors $v_1,v_2,v_3$ are linear independent and orthogonal to the vector $%
a|11>+b|12>+c|21>+d|22>$ in $H_A^2 \otimes H_B^2$.\\

$$
\begin{array}{ccccccccccc}
v_1=-c|11>+a|21> &  &  &  &  &  &  &  &  &  &  \\ 
v_2=-d|12>+b|22> &  &  &  &  &  &  &  &  &  &  \\ 
v_3=-(c+d)|11>-(c+d)|12>+(a+b)|21>+(a+b)|22> &  &  &  &  &  &  &  &  &  & 
\end{array}
$$

We have a family of dimension 3 subspaces span by $v_1,v_2,v_3$ in $H_A^2
\otimes H_B^2$. Since any dimension 3 subspace of $H_A^2 \otimes H_B^2$ has
only one normal direction, thus this example showed that generic dimension 3
subspaces (Here generic dimension 3 subspaces means that their normal
directions $a|11>+b|12>+c|21>+d|22>$ are outside the algebraic set defined
by $d=0$ or $ad=bc$.) of $H_A^2 \otimes H_B^2$ are linear span of seaprable
pure states. In high rank mixed state entanglement , it seems people cannot
only use ranges to determine whether mixed states are entangled and
eigenvalues play certain role, as partially manisfested in the result in [13] for highest rank mixed states.\\

In conclusion, we have proposed a separability criterion based on algebraic-geometric invariants of bipartite mixed states in $H_A^m \otimes H_B^n$ and proved that it can be used to detect all rank $r \leq m+n-3$ enatngled mixed states outside a measure zero set. Moreover it is proved that for many cases of low ranks $r \leq m+n-3 $ ,generic 
rank $r$ mixed states in $H_A^m \otimes H_B^m$ have relatively high Schmidt
numbers and thus highly entangled.  Our method can be used
constrcutively to get strong lower bounds of Schmidt numbers of low rank
bipartite mixed states as shown in Examples We also presented an Example to show that higher rank mixed state entanglement is quite different with low rank cases.\\

The author acknowledges the support from NNSF China, Information Science
Division, grant 69972049.\\

e-mail: chenhao1964cn@yahoo.com.cn\newline

\begin{center}
REFERENCES
\end{center}

1.Hao Chen, quant-ph/01008093\\

2.A.Einstein, B.Podolsky and N.Rosen, Phys. Rev. 47,777(1935)\\

3.E.Schrodinger, Proc.Camb.Philos.Soc.,31,555(1935)\\

4.R.Jozsa, in The Geometric Universe, edited by S.Huggett, L.Mason, K.P.Tod,
S.T.Tsou, and N.M.J.Woodhouse (Oxford Univ. Press, 1997)\\

5.A.Ekert, Phys Rev.Lett.67, 661(1991)\\

6.C.H.Bennett, G.Brassard, C.Crepeau, R.Jozsa, A.Peres and W.K.Wootters,
Phys.Rev.Lett 70, 1895 (1993)\\

7.C.H.Bennett, G.Brassard, S.Popescu, B.Schumacher, J.Smolin and
W.K.Wootters, Phys. Rev.Lett. 76, 722(1996)\\

8.M.Horodecki, P.Horodecki and R.Horodecki, in Quantum Information--Basic
concepts and experiments, edited by G.Adler and M.Wiener (Springer Berlin,
2000)\\

9.P.Horodecki and R.Horodecki, Quantum Information and Computation, Vol.1, no.1(2000)
,45-75\\

10.M.A.Nielsen and J.Kempe, Phys. Rev. Lett., 86, 5184(2001)\\

11.A.Peres, Phys. Rev. Lett., 77, 1413(1996)\\

12.M.Horodecki and P.Horodecki, Phys. Rev. A, 59, 4026(1999); N.Cerf, C.Adani and
R.M.Gingrich, Phys. Rev. A 60 898(1999)\\

13.P.Horodecki, M.Lewenstein, G.Vidal and I.Cirac, Phys. Rev. A 62, 032302 (2000)\\

14.P.Horodecki, Phys Lett A 232 (1997)\\

15.M.Horodecki, P.Horodecki and R.Horodecki, Phys.Rev.Lett. 80, 5239(1998)\\

16.C.H.Bennett, D.P.DiVincenzo, T.Mor,P.W.Shor, J.A.Smolin and T.M. Terhal, Phys Rev. Lett. 82, 5385 (1999)\\

17.D.DiVincenzo, T.Mor, P.Shor, J.Smolin and B.M.Terhal, Commun. Math. Phys. (accepted) quant-ph/9908070\\

18.D.Bruss and  A.Peres, Phys. Rev. A 61 30301(R) (2000)\\

19.P.Horodecki and M.Lewenstein, Phys. Rev. Lett. 85, 2657 (2000)\\

20.A.C.Doherty, P.A.Parrilo and F.M.Spedalier, Phys. Rev. Lett., 88, 187904(2002)\\

21.O.Rudolph, quant-ph/0202121\\

22.K.Chen, L-A.Wu and L.Yang, quant-ph/0205017\\

23.J.Harris, Algebraic Geometry, A First course, GTM 133, Springer-Verlag
1992\\

24.B.M.Terhal and P.Horodecki, Phys. Rev. A R040301, 61(2000)\\

25.S.Popescu,Phys.Rev.Lett. 72,797(1994),74,2619(1995)\\

26.K.Zyczkowski,P.Horodecki, A.Sanpera and M.Lewenstein, Phys.Rev. A, 58, 883(1998)\\

27.E.Arbarello, M.Cornalba, P.A.Griffiths and J.Harris, Geometry of
algebraic curves,Volume I, Springer-Verlag, 1985, Chapter II''Determinantal
Varieties''\\

\end{document}